\documentclass[twocolumn,amsmath,amssymb]{revtex4}

\def\lsim{\raise0.3ex\hbox{$\;<$\kern-0.75em\raise-1.1ex\hbox{$\sim\;$}}}
\def\gsim{\raise0.3ex\hbox{$\;>$\kern-0.75em\raise-1.1ex\hbox{$\sim\;$}}}

\usepackage{graphics}
\def\met{\slashed E_T}
\usepackage{epsfig}
\usepackage{slashed}
\usepackage{multirow}
\usepackage{pstricks}
\usepackage{dcolumn}
\newcommand{\be}{\begin{eqnarray}}
\newcommand{\ee}{\end{eqnarray}}

\def\bea{\begin{eqnarray}}
\def\eea{\end{eqnarray}}

\usepackage{epsfig}
\usepackage{array}
\usepackage{booktabs}
\usepackage{slashed}

\begin{document}
\title{$Z'$-induced Invisible Right-handed Sneutrino Decays at the LHC}
\author{W. Abdallah$^{1,2}$, J. Fiaschi$^3$, S. Khalil$^{1,4}$ and S. Moretti$^{3}$ }
\vspace*{0.2cm}
\affiliation{$^1$Center for Fundamental Physics, Zewail City of Science and Technology, 6 October City, Giza 12588, Egypt.\\
$^2$Department of Mathematics, Faculty of Science, Cairo
University, Giza 12613, Egypt.\\
$^3$School of Physics and Astronomy, University of Southampton,
Highfield, Southampton SO17 1BJ, UK.\\
$^4$Department of Mathematics, Faculty of
Science, Ain Shams University, Cairo 11566, Egypt.}
\date{\today}

\begin{abstract}

The invisible signals of right-handed sneutrino decays originating from a $Z'$ 
are analyzed at the Large Hadron Collider. The possibility of accessing 
these events  helps disentangling the $B-L$ extension of Minimal Supersymmetric Standard Model from more popular scenarios of Supersymmetry. We assess the scope of the CERN machine in establishing the aforementioned signatures when accompanied by mono-jet, single-photon or $Z$-radiation probes through sophisticated signal-to-background simulations carried out in presence of parton shower, hadronisation as well as detector effects.  We find substantial sensitivity to all such signals for standard luminosities at Run 2.

\end{abstract}
\maketitle
The $B-L$ Supersymmetric Standard Model (BLSSM) represents an appealing non-minimal realization of 
Supersymmetry (SUSY), as it is more compatible with current Large Hadron Collider (LHC) data than the Minimal 
Supersymmetric Standard Model (MSSM) and further uniquely  accounts for the well established existence of non-zero neutrino masses. In this scenario, which is based on the gauge group $SU(3)_C \times SU(2)_L \times U(1)_Y \times U(1)_{B-L}$, (heavy) right-handed neutrino Superfields are introduced in order to implement a Type I seesaw~\cite{ref:BL}. Also, it has been shown that the scale of the $B-L$ symmetry breaking is related to the soft SUSY scale~\cite{Khalil:2007dr}. Furthermore, the BLSSM also alleviates the 
little hierarchy problem of the MSSM, as both the additional singlet Higgs state and right-handed (s)neutrinos release additional parameter space from the LEP, Tevatron and LHC constraints.

The $Z'$ and (s)neutrino sectors are ideal hallmark manifestations of the BLSSM as candidate underlying SUSY model. An intriguing 
signal would be the one involving totally invisible decays of a $Z'$ into (s)neutrinos, thereby being
potentially accessible in mono-jet, single-photon and $Z$-ISR
(Initial State Radiation) analyses. Contrary to SUSY models which do not have a $Z'$ in their spectra, in the BLSSM one can afford resonant $Z'$  (with a TeV scale mass) production and decay into heavy (s)neutrinos which can in turn decay, again on-shell, into an invisible final state. Therefore, one would expect the typical BLSSM distributions of the visible probe (whether it be mono-jet, single-photon or $Z$-ISR) to be substantially different from other SUSY
scenarios where the mediator is much lighter. In this letter, we aim at studying this phenomenology and assess the scope of the LHC in testing it. It is worth mentioning that visible decays of the $Z'$ into (s)neutrinos at the LHC have been studied in~\cite{Elsayed:2012ec}.

The particle content of the  BLSSM includes, in addition to the MSSM fields, three chiral right-handed Superfields ($N_i$), a vector Superfield associated
to $U(1)_{B-L}$ ($Z'$) and two chiral Higgs singlet Superfields ($\chi_1$, $\chi_2$).  After $B-L$ symmetry breaking has taken place, the $U(1)_{B-L}$ gauge boson acquires a mass~\cite{ref:BL}, $M^2_{Z'}=g^2_{B-L}v'^2$, which is function of the $B-L$ gauge coupling and the Higgs singlet vacuum expectation value. The most stringent constraint on the
$U(1)_{B-L}$ gauge boson parameters is obtained from LEP2 results, which imply $\frac{M_{Z'}}{g_{B-L}}>6$~TeV~\cite{Cacciapaglia:2006pk}.
However, one should note that this result is based on the assumption that  the $Z'$ dominantly decays to the Standard Model
(SM) quarks and leptons and  is derived from the limit on the low energy four-fermion contact interactions induced by the $Z'$ exchange diagram.  As such, this type of bound cannot per se be relaxed, as one cannot account for, e.g., 
 invisible decays of the $Z'$ (into right-handed (s)neutrinos, in the BLSSM). In fact, if one departs from the effective Lagrangian
approach, other dynamics may come into play, e.g., non-negligible interference effects between the $Z'$ and the SM neutral gauge bosons (again, as typical in the BLSSM). All this being said, since a re-computation of the LEP limits in the BLSSM is beyond the scope of this work, we adopt the 6 TeV limit as conservative in our scenario and nonetheless use it to constrain the $g_{B-L}$ coupling~\cite{Carena}, which should then be $g_{B-L} < M_{Z'} / 6$ TeV.

The $Z'$ parameters should also be consistent with  the exclusion limits imposed by the LHC Run 1. The latter are obtained from the Drell-Yan neutral channel as a function of the dilepton Branching Ratio (BR) of the $Z'$, which in the BLSSM
depends on a variety of parameters. However, rather than scanning on the latter (which will be the scope of a separate
publication~\cite{progress}) we have selected here two representative configurations of the BLSSM, one with a narrow and the other with a wide resonance, namely, $\Gamma_{Z'} / M_{Z'} \simeq 2 \%$ and $\simeq 40 \%$, respectively. In case of  a narrow resonance,  one finds that $M_{Z'} \gsim 2.5$~TeV whereas, for the wider resonance, this bound can be lessened to ${\cal O}(1~{\rm TeV})$, both for $g_{B-L}$ couplings that  
we have verified to be not excluded by the data collected at the LHC with  
a center of mass energy of 8 TeV and an integrated luminosity of 20 fb$^{-1}$.
In order to perform such LHC tests, we have first reproduced experimental searches appropriate for both the scenarios of narrow and wide resonance in the dilepton final states (with both electrons and muons).
In both cases we have included Next-to-Next-to-Leading Order 
(NNLO) QCD corrections as well as the quoted acceptances and efficiencies in~\cite{Khachatryan:2014fba} for the 
electron and muon channels. We have finally combined the signal and background rates obtained from the two 
leptonic signatures and verified that our benchmarks have a significance below 2.
In the narrow case we integrated the signal in a small mass region around the peak, chosen in order to maximize the respective significance, the latter 
obtained using Poisson statistics.
In the wide case, since the peak structure in not visible, we have adopted a ``counting experiment'' approach. We have integrated the new physics 
signal imposing a low cut in the invariant mass distribution and have verified that the excess of $Z'$ events (above the SM ones) is not sufficient to claim a 
discovery, adopting Gaussian statistics. Similarly, we did with regard to the bounds in~\cite{Aad:2014cka}. 
In the remainder we will focus on a narrow  $Z'$, for sake of illustration. Further, for our chosen benchmark, all
sparticle states, especially the colored ones, are heavy (but the sneutrinos and some of the charginos/neutralinos entering the $Z'$ decay chains
yielding invisible signals), so that no sizable intrinsic SUSY background exists.

\begin{figure*}[t]
\includegraphics[width=8.5cm,height=6.5cm]{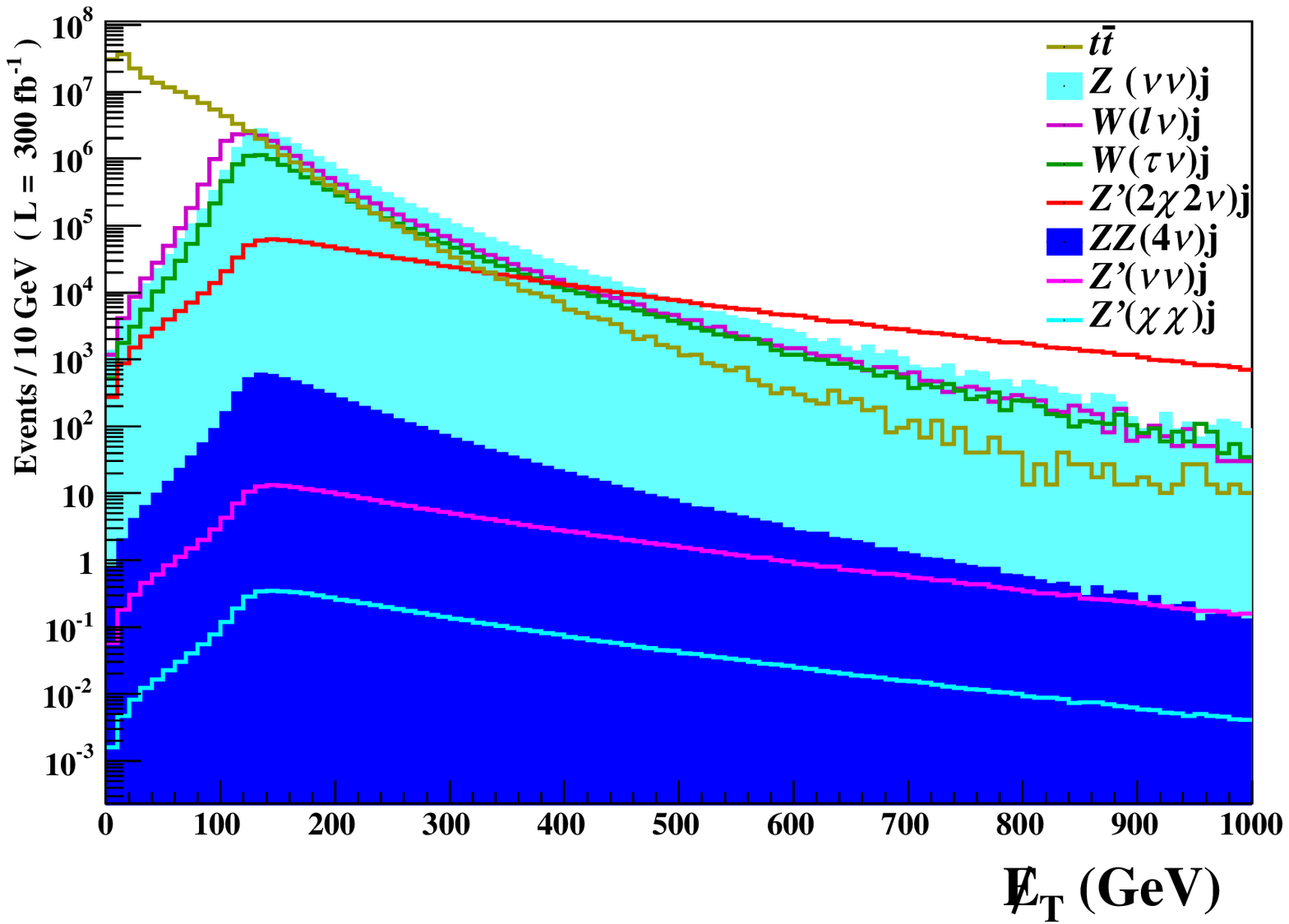} 
\includegraphics[width=8.5cm,height=6.5cm]{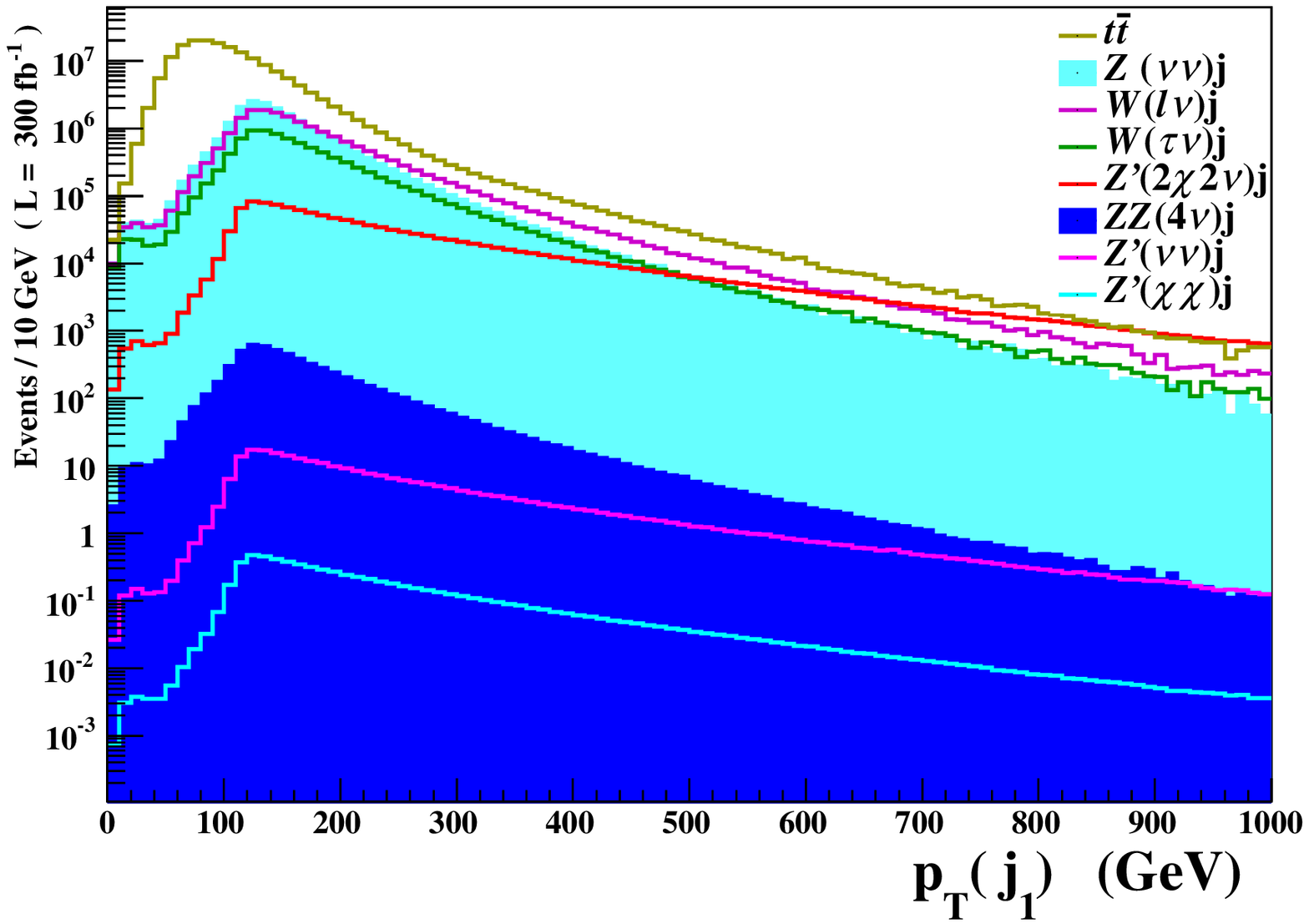} 
\caption{(Left panel) Number of events versus the missing transverse energy. (Right panel) Number of events versus the transverse  momentum of the leading jet. Distributions are for the mono-jet case given after the jet selection only.
The energy is $14$~TeV whereas the integrated luminosity is 300~fb$^{-1}$. 
Here, $M_{Z'}=2.5$~TeV and $g_{B-L}=0.4$.}
\label{fig1}
\end{figure*}

\begin{table*}[t]
{\small\fontsize{8.5}{8.5}\selectfont{
\begin{tabular}{|c||c|c|c|c|c|c|c|c|c|}
\hline
\multicolumn{2}{|c|}{}&\multicolumn{5}{c|}{Backgrounds}&\multicolumn{3}{c|}{Signals}\\
\hline
\multicolumn{2}{|c|}{ Process}& $Z(\nu\bar{\nu})j$ & $W(l\nu_l)j$ & $W(\tau\nu_\tau)j$ &$t\bar{t}$& $ZZj$  
&$Z'(2\tilde\chi \,2\nu)j$&$Z'(\nu\bar{\nu})j$&$Z'(\tilde\chi\tilde\chi)j$ \\
\hline
\hline
\multicolumn{2}{|c|}{Before cuts} &21573000 & 19248000 & 9390000 & 179058000 & 6621 & 1334400 & 278 & 7.54  \\
\hline
\multirow{9}{*}{\rotatebox{90}{Cut}}&(1) &16823567 \!$\pm$\! 1924& 15817945 \!$\pm$\! 1678& 7719914 \!$\pm$\! 1171& 151390826 \!$\pm$\! 4836& 5732 \!$\pm$\! 28& 1219314 \!$\pm$\! 324& 255 \!$\pm$\! 4.68& 6.895 \!$\pm$\! 0.77 \\
\cline{2-10}
&(2)&65275 \!$\pm$\! 255& 135191 \!$\pm$\! 366& 65423 \!$\pm$\! 254& 298430 \!$\pm$\! 545& 73 \!$\pm$\! 8.5& 130636 \!$\pm$\! 343& 27 \!$\pm$\! 4.95& 0.741 \!$\pm$\! 0.82  \\
\cline{2-10}
&(3)&45530 \!$\pm$\! 213& 32569 \!$\pm$\! 180& 27102 \!$\pm$\! 164& 6836.8 \!$\pm$\! 82.7& 55.6 \!$\pm$\! 7.43& 118456 \!$\pm$\! 328& 25 \!$\pm$\! 4.74& 0.672 \!$\pm$\! 0.78  \\
\cline{2-10}
&(4)&14283 \!$\pm$\! 119& 10566 \!$\pm$\! 102& 8668.5 \!$\pm$\! 93.1& 2808 \!$\pm$\! 53& 16.5 \!$\pm$\! 4.06& 35424 \!$\pm$\! 185& 7.4 \!$\pm$\! 2.68& 0.201 \!$\pm$\! 0.44 \\
\cline{2-10}
&(5)&10831 \!$\pm$\! 104& 7395.3 \!$\pm$\! 86& 6088.7 \!$\pm$\! 78& 881.7 \!$\pm$\! 29.7& 12.2 \!$\pm$\! 3.49& 23330 \!$\pm$\! 151& 4.9 \!$\pm$\! 2.18& 0.132 \!$\pm$\! 0.36  \\
\cline{2-10}
&(6)&8992.5 \!$\pm$\! 94.8& 6007.4 \!$\pm$\! 77.5& 4699.9 \!$\pm$\! 68.5& 379.8 \!$\pm$\! 19.5& 9.79 \!$\pm$\! 3.13& 18806 \!$\pm$\! 136& 3.9 \!$\pm$\! 1.96& 0.107 \!$\pm$\! 0.33  \\
\cline{2-10}
&(7)&8969.8 \!$\pm$\! 94.7& 3343.1 \!$\pm$\! 57.8& 3929 \!$\pm$\! 62.7& 257.7 \!$\pm$\! 16.1& 9.78 \!$\pm$\! 3.12& 18786 \!$\pm$\! 136& 3.9 \!$\pm$\! 1.96& 0.107 \!$\pm$\! 0.32  \\
\cline{2-10}
&(8)&8969.8 \!$\pm$\! 94.7& 871.2 \!$\pm$\! 29.5& 3207.4 \!$\pm$\! 56.6& 176.3 \!$\pm$\! 13.3& 9.77 \!$\pm$\! 3.12& 18782 \!$\pm$\! 136& 3.9 \!$\pm$\! 1.96& 0.107 \!$\pm$\! 0.32  \\
\cline{2-10}
&(9)&8458.9 \!$\pm$\! 92& 790.2 \!$\pm$\! 28.1& 1378.8 \!$\pm$\! 37.1& 81.39 \!$\pm$\! 9.02& 9.21 \!$\pm$\! 3.03& 17878 \!$\pm$\! 132& 3.7 \!$\pm$\! 1.92& 0.102 \!$\pm$\! 0.32  \\
\cline{2-10}
&(10)&8152.3 \!$\pm$\! 90.3& 769.9 \!$\pm$\! 27.7& 1334.4 \!$\pm$\! 36.5& 54.26 \!$\pm$\! 7.37& 8.8 \!$\pm$\! 2.96& 17357 \!$\pm$\! 130& 3.6 \!$\pm$\! 1.89& 0.098 \!$\pm$\! 0.31  \\
\hline
\end{tabular}}}
\caption{The cut flow on background versus signal events for $M_{Z'}=2.5$~TeV and $g_{B-L}=0.4$ in the mono-jet channel at $\sqrt s=14$~TeV with 
${\cal L}dt= 300$~fb$^{-1}$: (1) $n(\text{jets})\geq 1$ with $|\eta(j_1)|<2$; (2) $p_T(j_1)> 500$~GeV; (3) $\met > 500$~GeV; (4) $\Delta \phi(j_2,\met) > 0.5$; (5) veto on $p_T(j_2)> 100$~GeV, $|\eta(j_2)|< 2$; (6) veto on $p_T(j_3)> 30$~GeV, $|\eta(j_3)|< 4.5$; (7) veto on e; (8) veto on $\mu$; (9) veto on $\tau$-jets; (10) veto on $b$-jets.}
\label{tab1} 
\end{table*}

From the BLSSM Lagrangian, the relevant interactions for the right-handed sneutrinos are given by
\begin{eqnarray}
{\cal L}_{int}^{^{\tilde{\nu}_R}} \!&\!=\!&\! (Y_\nu)_{ij} \bar{l}_i P_R (V_{k2} \tilde{\chi}^+_k)^\dag (\Gamma_{\nu_R})_{\alpha j} \tilde{\nu}_{R_\alpha} \\
\!&\!+\!&\! (Y_{\nu})_{ij}(U_{\rm MNS})_{il} \bar{\nu}_l P_R (N_{k1}^* \tilde{\chi}_k^0) (\Gamma_{\nu_R})_{j\alpha} \tilde{\nu}_{R_\alpha}  \nonumber\\
\!&\!+\!&\! (Y_\nu)_{ij} (M_N)_j \cos \beta (\Gamma_{L_L})_{\beta i} \tilde{l}_\beta H^+ (\Gamma_{\nu_R})_{\alpha j}\tilde{\nu}_{R_\alpha} .\nonumber
\end{eqnarray}
Here, the rotational matrices $\Gamma_{L_L}$ and $\Gamma_{\nu_R}$ are defined as $\Gamma_L \equiv (\Gamma_{L_L}, \Gamma_{L_R})$ and $\Gamma_\nu \equiv (\Gamma_{\nu_L}, \Gamma_{\nu_R})$. From these interactions, it can be easily concluded that, if the lightest right-handed sneutrino is lighter than the lightest slepton and lightest chargino, then it decays into light SM-like neutrinos and lightest neutralinos. This decay channel would be invisible, since both light neutrinos and lightest
neutralinos would be escaping the detector. Hence, given the discussed SUSY construction peculiar to the BLSSM,
this pattern can provide  a robust signature of $B-L$ sneutrinos through 
 mono-jet, single-photon  and/or $Z$-ISR, further considering that the high $Z'$ mass will force the missing transverse
energy (and hence the $p_T$ of the visible probe) to be harder than in $Z$ (into neutralinos) mediated events typical of the MSSM.

In our calculations we have used SARAH~\cite{florianSARAH} and SPheno~\cite{PorodSPheno,florianSPheno} to build the BLSSM and calculate masses, couplings and Branching Rations (BRs).  We also considered the following benchmark (given at the SUSY scale): $M_{Z'} = 2.5$~TeV, $g_{B-L} = 0.4$, $M_{\tilde{\nu}_{R_1}} \simeq M_{\tilde{\nu}_{R_2}}\simeq M_{\tilde{\nu}_{R_3}} \simeq 580$~GeV,  $M_{\tilde{\nu}_{R_4}} \simeq M_{\tilde{\nu}_{R_5}}\simeq M_{\tilde{\nu}_{R_6}} \simeq 740$~GeV, $m_{\tilde{\chi}^\pm_{1,2}}\simeq 4, 0.9$~TeV, $m_{\tilde{\chi}^0_{1}}\simeq 440$~GeV and slepton masses  of order $700$~GeV. Furthermore, the matrix-elements  for the  parton level signals and backgrounds were derived from MadGraph5~\cite{Madgraph5}. Then, for showering and hadronisation we have used PYTHIA~\cite{pythia} whereas  we have performed the fast detector simulations with PGS4~\cite{PGS4}. As jet finding algorithm, we have used a cone (with calorimeter $k_t$ cluster finder) of size $\Delta R = 0.5$. Then, we have manipulated the generated data with MadAnalysis5~\cite{Madanalysis5}. Finally,
we have adopted  usual selection strategies, wherein cuts are enforced against the kinematics of the highest $p_T$ jet/photon or the $Z$. 

We start with the mono-jet case, which is generally dominated by sneutrino decays \footnote{Also the
$Z'\to\nu\bar{\nu}$ and $\tilde{\chi}^{0}_{1}\tilde{\chi}^{0}_{1}$ 
invisible decays are present, but they are subleading in our benchmark.} (hereafter, $j=$jet), 
\begin{eqnarray}
q\bar{q}\to Z'(\to\tilde\nu_R\tilde\nu^*_R\to \tilde\chi^{0}_{1}\tilde\chi^{0}_{1}\,\nu\bar\nu)+j.
\label{monojetsignal}
\end{eqnarray}
The SM backgrounds are the following: $ Z(\to \nu\bar{\nu})+j$ (irreducible) plus $W(\to l \nu)+j$, $W(\to \tau \nu)+j$, $ t\bar{t}$ and $ZZ(\to 2\nu 2\bar{\nu}) +j$ (all reducible). We closely follow here the selection of~\cite{Baer, Saavedra}. Further, in order to increase the Monte Carlo 
efficiency and thus obtain reasonable statistics, we have applied a parton level cut of $p_T(j_1) > 120$~GeV for both signals and backgrounds (here $p_T(j_1)$ is the highest jet transverse momentum). According to the estimation of the QCD background based on the full detector simulation of~\cite{multijet,allanach}, such a noise can be reduced to a negligible level by requiring a large $\slashed E_T$ cut. Thanks to the heavy $Z'$ mediation, we can afford to set here $\slashed E_T > 500$~GeV~\cite{drees}. In view of this we can then also implement a $\slashed E_T > 100$~GeV cut for both signals and backgrounds at generation level. 
The beneficial effect of the $\slashed E_T$ selection is evident from the left plot in Fig.~\ref{fig1}.  In contrast, a similar cut on $p_T(j_1)$ is not as selective (and is  anyway correlated),  see the right plot in Fig.~\ref{fig1},
yet it pays off to also enforce it.

In Tab.~\ref{tab1}, we present the actual cut flow for signal and background events, given
 at 14~TeV with an integrated luminosity of 300~fb$^{-1}$. After the BLSSM specific cuts, i.e.,  $\slashed E_T$ and 
$p_T(j_1)>500$~GeV, all the backgrounds are reduced under the dominant sneutrino signal. Other key steps of the 
analysis are the lepton, $\tau$- and $b$-jet vetoes, which suppress the $W+j$ and $t\bar{t}$ backgrounds by more than two orders of magnitude~\cite{lhc}.

\begin{figure*}[t]
\includegraphics[width=8.5cm,height=7cm]{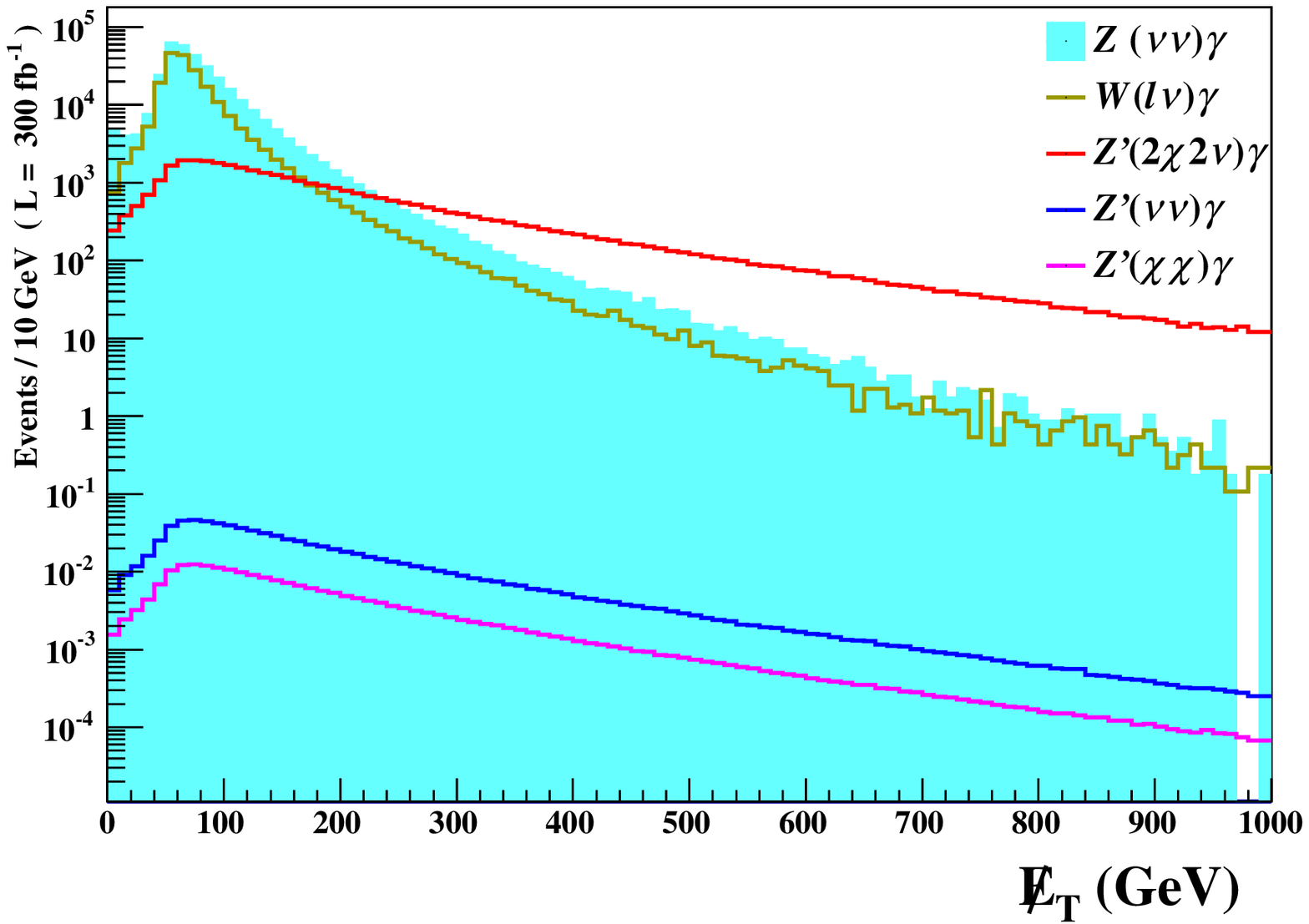} 
\includegraphics[width=8.5cm,height=7cm]{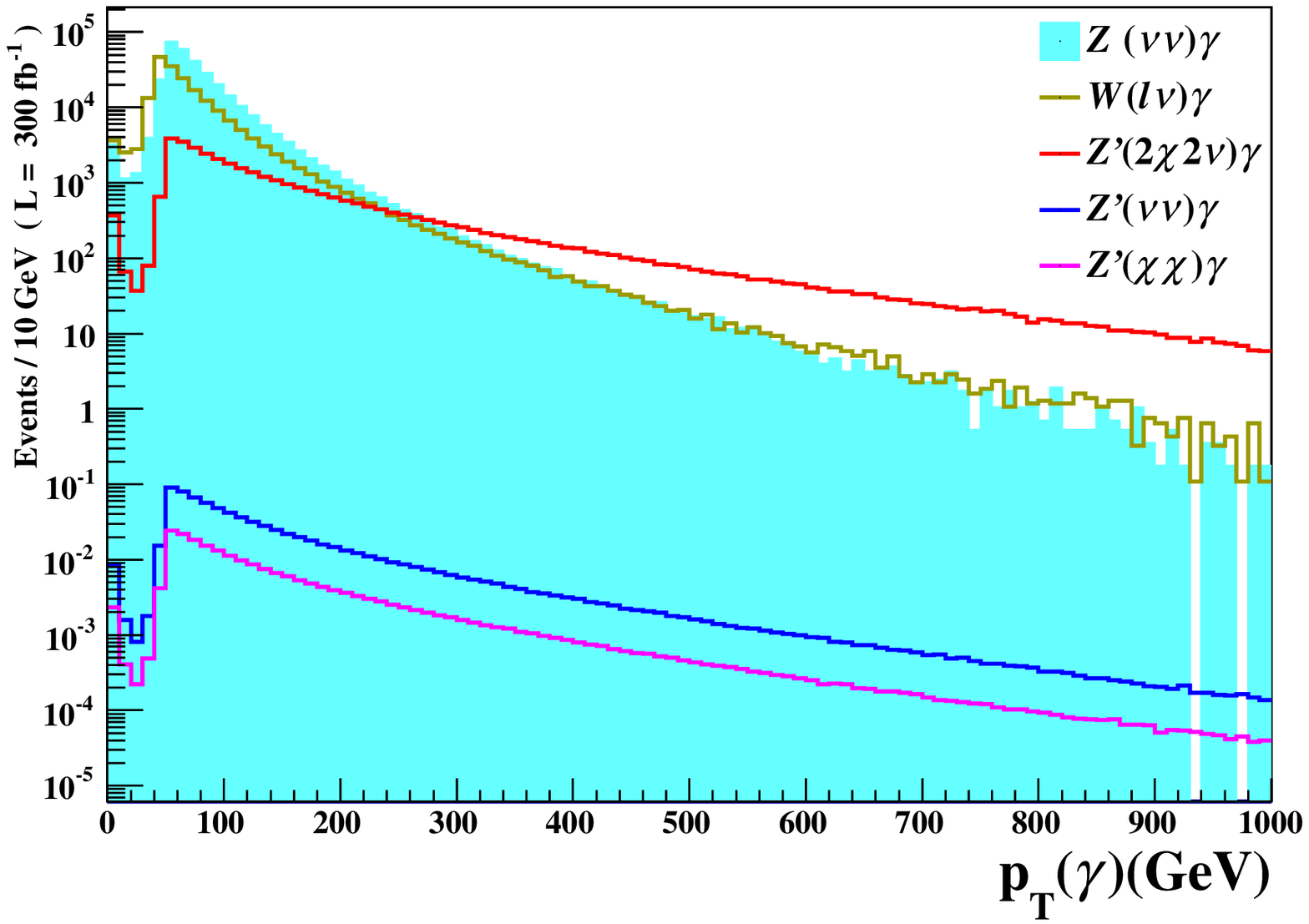}
\caption{(Left panel) Number of events versus the missing transverse energy. (Right panel) Number of events versus the transverse  momentum of the leading photon. Distributions are for the single-photon case given after the  jet selection only.
The energy is 14 TeV whereas the integrated luminosity is 300~fb$^{-1}$. 
Here, $M_{Z'}=2.5$~TeV and $g_{B-L}=0.4$.}
\label{fig3}
\end{figure*}

Now we turn to the single-photon signature, which occurs in our
BLSSM point mainly via the following process: 
\begin{eqnarray}
q\bar{q}\to Z' (\to \tilde{\nu}_R \tilde{\nu}_R^* \to \tilde{\chi}_1^0 \tilde{\chi}_1^0 \nu \bar{\nu}) + \gamma. 
\end{eqnarray}
We have generated single-photon events while requiring 
the following parton level (generation) cuts: $\met > 50$~GeV, $p_T(\gamma_1) > 40$~GeV, $p_T(j_1)> 25$~GeV
($p_T(\gamma_1)$ being the leading photon transverse momentum). We also generate the background processes $Z(\to\nu\bar{\nu})+\gamma$ and $W(\to l\nu_{l})+\gamma$, where $l =e$,
$\mu$ or $\tau$. The latter are reduced significantly by applying the cut flow shown in Tab.~\ref{tab:monophoton1}, see~\cite{Baer, atlas} for guidance. Again, the sneutrino signal emerges dramatically over the backgrounds.
Further, in Fig.~\ref{fig3},  the spectra in  $\met$ and $p_T(\gamma_1 )$ are  
 shown. These plots well motivate our high transverse energy/momentum cuts, however, we note that the smaller signal rates here force us to a softer, yet still very effective, requirement, of a $150$~GeV threshold.

\begin{table*}
{\small\fontsize{9}{9}\selectfont{
\begin{tabular}{|c||c|c|c|c|c|c|}
\hline
\multicolumn{2}{|c|}{}&\multicolumn{2}{c|}{Backgrounds}&\multicolumn{3}{c|}{Signals}\\
\hline
\multicolumn{2}{|c|}{ Process}& $Z(\nu\bar{\nu})\gamma$ & $W(l\nu_l)\gamma$&$Z'(2\tilde\chi \,2\nu)\gamma$&$Z'(\nu\bar{\nu})\gamma$&$Z'(\tilde\chi\tilde\chi)\gamma$ \\
\hline
\hline
\multicolumn{2}{|c|}{Before cuts}& 332712 & 204644 & 37380 & 0.861 & 0.234  \\
\hline
\multirow{6}{*}{\rotatebox{90}{Cut}}&$n(\gamma)\geq 1$  & 316031 \!$\pm$\! 125& 192677 \!$\pm$\! 106& 34998 \!$\pm$\! 47.2& 0.806 \!$\pm$\! 0.227& 0.219 \!$\pm$\! 0.118 \\
\cline{2-7}
&$p_T(\gamma_1)> 150$~GeV  & 18576 \!$\pm$\! 132& 12146 \!$\pm$\! 106& 12357.8 \!$\pm$\! 91& 0.282 \!$\pm$\! 0.435& 0.0765 \!$\pm$\! 0.2268  \\
\cline{2-7}
&$\met > 150$~GeV  & 14681 \!$\pm$\! 118& 4287.3 \!$\pm$\! 64.8& 11202 \!$\pm$\! 88.6& 0.255 \!$\pm$\! 0.424& 0.0693 \!$\pm$\! 0.2208  \\
\cline{2-7}
&$n(j)\leq 1$, $|\eta(j)|< 4.5$  & 6819.7 \!$\pm$\! 81.7& 2388.2 \!$\pm$\! 48.6& 7415 \!$\pm$\! 77.1& 0.168 \!$\pm$\! 0.368& 0.0457 \!$\pm$\! 0.1917 \\
\cline{2-7}
&veto on e  & 6817.6 \!$\pm$\! 81.7& 1731.8 \!$\pm$\! 41.4& 7409.3 \!$\pm$\! 77.1& 0.168 \!$\pm$\! 0.368& 0.0456 \!$\pm$\! 0.1916  \\
\cline{2-7}
&veto on $\mu$  & 6817.6 \!$\pm$\! 81.7& 1132.5 \!$\pm$\! 33.6& 7407 \!$\pm$\! 77.1& 0.168 \!$\pm$\! 0.368& 0.0456 \!$\pm$\! 0.1916  \\
\cline{2-7}
&veto on $\tau$-jets  & 6479.8 \!$\pm$\! 79.7& 758 \!$\pm$\! 27.5& 7069 \!$\pm$\! 75.7& 0.161 \!$\pm$\! 0.631& 0.0435 \!$\pm$\! 0.1882 \\
\hline
\end{tabular}}}
\caption{The cut flow on background versus signal events for $M_{Z'}=2.5$~TeV and $g_{B-L}=0.4$ in the single-photon channel at $\sqrt s=14$~TeV with 
${\cal L}dt= 300$~fb$^{-1}$.}
\label{tab:monophoton1}
\end{table*}

\begin{figure}[ht]
\begin{center}
\includegraphics[width=9cm,height=7cm]{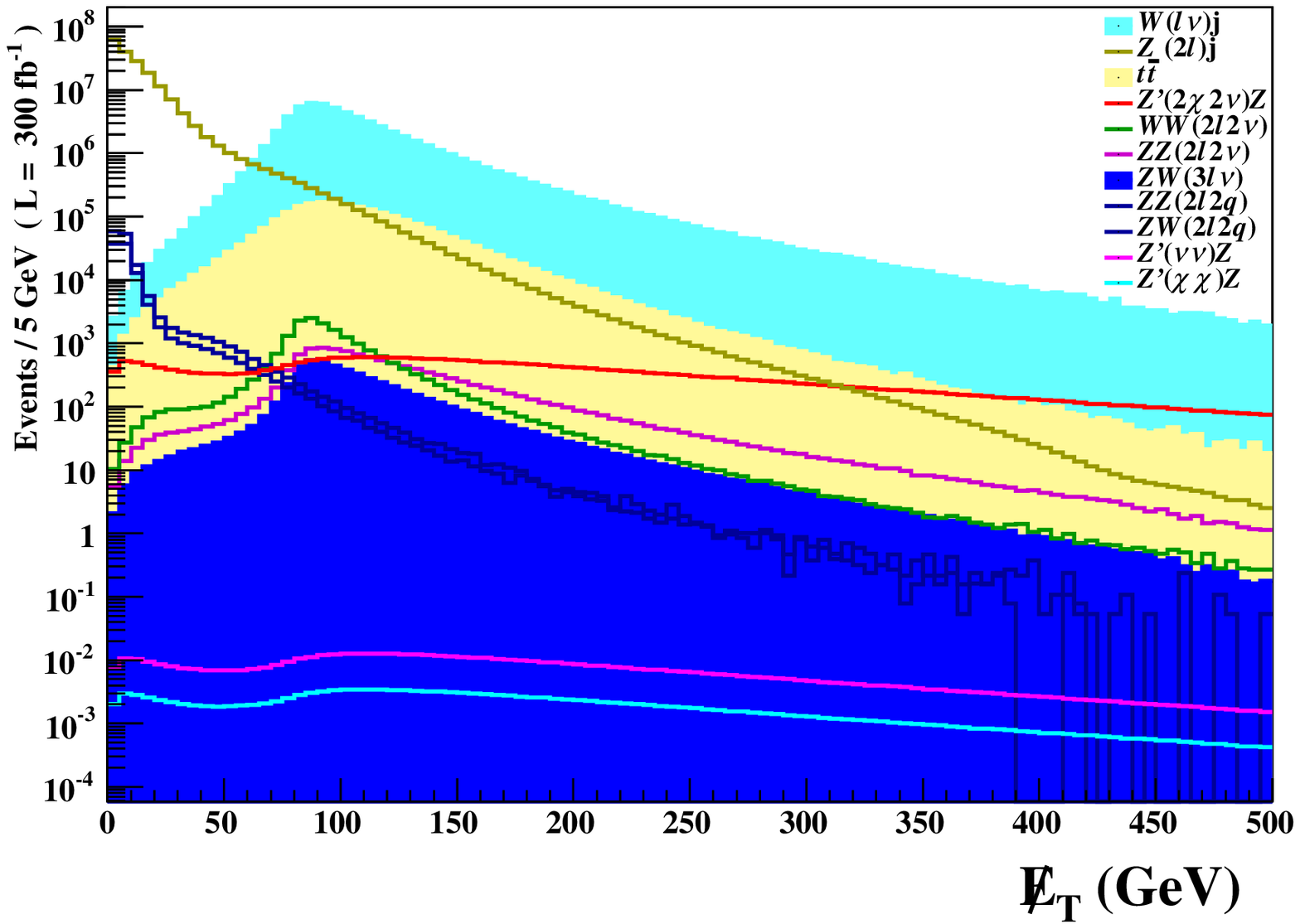} 
\end{center}
\caption{Number of events versus the missing transverse energy. Distributions are for the $Z$-ISR case given after the  jet selection only. The energy is $14$~TeV whereas the integrated luminosity is 300~fb$^{-1}$. 
Here, $M_{Z'}=2.5$~TeV and $g_{B-L}=0.4$.}
\label{fig5}
\end{figure}
For the $Z$-ISR signature in the BLSSM, 
\begin{eqnarray}
q\bar{q}\to Z' (\to \tilde{\nu}_R \tilde{\nu}_R^* \to \tilde{\chi}_1^0 \tilde{\chi}_1^0 \nu \bar{\nu}) + Z, 
\end{eqnarray}
we generate events with the following parton level cuts: $\slashed E_T>80$~GeV, $p_T(l)>10$~GeV and $p_T(j)>20$~GeV. The dominant irreducible background is $ZZ\to l^+ l^- \bar\nu\nu$  ($l = e,\mu$) and the other noise in this category is $WW\to l^+ \nu l^- \bar\nu $. As we reconstruct the $Z$ probe,
we enforce a cut on an invariant mass window centered on the $Z$ mass for two oppositely charged leptons, $m_{ll}\in [76,106]$~GeV~\cite{Aad:2014vka},  the latter is strongly reduced. The reducible backgrounds may have jets produced: $Z+\text{\rm jets}$, $ZZ\to \bar{q} q l^+ l^-$ and $ZW\to l^+ l^- \bar{q}q$. After the customary large $\met$ cut that the heavy $Z'$ allows us to enforce (here, $\met > 250$~GeV), this cumulative noise yields no event for the luminosity adopted. In addition, there are other reducible backgrounds with jets  that we have dealt with: $t\bar{t} \to l^+ \nu b l^- \bar\nu\bar{b}$ which is reduced by rejecting events containing at least one jet with $p_T>25$~GeV; $W+$jets which is reduced by the large $\met$ cut. The last leptonic background is $ZW\to l\nu l^+ l^-$, which is also eliminated by the cut $\met > 250$~GeV. The cut flow (modeled on~\cite{Bell:2012rg}) and individual responses of signals and
backgrounds are shown in Tab.~\ref{tab:monoZ}. In Fig.~\ref{fig5}, we show the various distributions in $\met$, again,
for the purpose of justifying our BLSSM specific $\met$ selection. Again, the invisible signal dominated by the sneutrinos and accompanied by the reconstructed $Z$ stands well above the backgrounds.

\begin{table*}
{\small\fontsize{7}{7}\selectfont{
\begin{tabular}{|c||c|c|c|c|c|c|c|c|c|}
\hline
\multicolumn{2}{|c|}{}&\multicolumn{5}{c|}{Backgrounds}&\multicolumn{3}{c|}{Signals}\\
\hline
\multicolumn{2}{|c|}{ Process} & $ZZ(2l2\nu)$ & $WW(2l2\nu)$ & $ZW(3l\nu)$ &$W(l\nu)j$& $t\bar{t}$  
&$Z'(2\tilde\chi \,2\nu)Z$&$Z'(\nu\bar{\nu})Z$&$Z'(\tilde\chi\tilde\chi)Z$ \\
\hline
\hline
\multicolumn{2}{|c|}{Before cuts} &12027 & 18966 & 5541 & 64980000 & 2377500 & 33900 & 0.703 & 0.191  \\
\hline
\multirow{3}{*}{\rotatebox{90}{Cut}}&$ m_{ll}\in [76,106]$~GeV &9068.1 \!$\pm$\! 47.2 & 2726.2 \!$\pm$\! 48.3 & 4392.8 \!$\pm$\! 30.2& 521652 \!$\pm$\! 719 & 403272 \!$\pm$\! 578 & 1553.3 \!$\pm$\! 38.5& 0.0322 \!$\pm$\! 0.175& 0.0088 \!$\pm$\! 0.0914\\
\cline{2-10}
&veto on $p_T(j)> 25$~GeV  &6510.6 \!$\pm$\! 54.6& 2025.7 \!$\pm$\! 42.5& 2997.1 \!$\pm$\! 37.1& 193982 \!$\pm$\! 439& 12007 \!$\pm$\! 109& 696.2 \!$\pm$\! 26.1& 0.0145 \!$\pm$\! 0.119& 0.0039 \!$\pm$\! 0.0617  \\
\cline{2-10}
&$\met > 250$~GeV  &229 \!$\pm$\! 15.0& 1.15 \!$\pm$\! 1.07& 49.63 \!$\pm$\! 7.01& 171 \!$\pm$\! 13.1& 8.76 \!$\pm$\! 2.96& 200.3 \!$\pm$\! 14.1& 0.0041 \!$\pm$\! 0.064& 0.0011 \!$\pm$\! 0.0334  \\
\hline
\end{tabular}}}
\caption{The cut flow on background versus signal events for $M_{Z'}=2.5$~TeV and $g_{B-L}=0.4$ in the $Z$-ISR channel at $\sqrt s=$ 14    TeV with 
${\cal L}dt= 300$~fb$^{-1}$.}
\label{tab:monoZ}
\end{table*}

In summary, we have proven the sensitivity that the LHC has in Run 2 with standard luminosity settings in probing invisible signals which emerge in the BLSSM from $Z'$ decays in presence of an associated jet, photon or $Z$-boson. For all such signatures, we were able, upon enforcing well established selection procedures for these topologies
supplemented by BLSSM specific cuts, to establish signals with significances well above the customary 
$5\sigma$ discovery limit. Indeed, this has been possible thanks to the fact that the BLSSM mediator of such invisible signals is a very heavy $Z'$, with mass in the TeV region, thereby transferring to its decay products large transverse momenta that can be generically exploited in all cases for background reduction. Furthermore, for all topologies considered, the dominant
component of the signal is via sneutrinos (above neutrinos and neutralinos), so that assessing these invisible signatures in the heavy $\met$ regime would not only signal the presence of a dark matter induced channel within SUSY but also be a 
circumstantial evidence of a theoretically well motivated non-minimal version of it, the BLSSM. In a forthcoming publication~\cite{progress}, we shall show that the case made here for illustrative purposes using a benchmark with a 
2.5 TeV and rather narrow $Z'$ can be extended to a large  BLSSM parameter space volume (also covering lighter and/or wider $Z'$s). 

\section*{Acknowledgements}
The work of W.A. and S.K. is partially funded by the ICTP grant AC-80. J.F. and
S.M. are supported in part through the NExT Institute. The work of S.K. and S.M. is also funded through the
grant H2020-MSCA-RISE-2014 no. 645722 (NonMinimalHiggs). W.A. would like to thank M. Ashry, A. Ali and A. Moursy for fruitful discussions.



\begin{thebibliography}{99}
%
\bibitem{ref:BL}
  S.~Khalil,
  J.\ Phys.\ G {\bf 35}, 055001 (2008);
  L.~Basso, A.~Belyaev, S.~Moretti and C.~H.~Shepherd-Themistocleous,
  Phys.\ Rev.\ D {\bf 80}, 055030 (2009);
  S.~Iso, N.~Okada and Y.~Orikasa,
  Phys.\ Lett.\ B {\bf 676}, 81 (2009).
\bibitem{Khalil:2007dr}
  S.~Khalil and A.~Masiero,
  Phys. Lett. B {\bf 665}, 374 (2008);
  Z.~M.~Burell and N.~Okada,
  Phys. Rev. D {\bf 85}, 055011 (2012);
  P.~Fileviez Perez and S.~Spinner,
  Phys. Rev. D {\bf 83}, 035004 (2011).
\bibitem{Elsayed:2012ec} 
  A.~Elsayed, S.~Khalil, S.~Moretti and A.~Moursy,
  Phys. Rev. D {\bf 87}, 053010 (2013).
\bibitem{Cacciapaglia:2006pk}
G.~Cacciapaglia, C.~Csaki, G.~Marandella and A.~Strumia,
  Phys. Rev. D {\bf 74}, 033011 (2006).
\bibitem{Carena}
M. Carena, A. Daleo, B. A. Dobrescu and T. M. P. Tait, Phys. Rev. D {\bf 70}, 093009 (2004).  
\bibitem{progress}
W.~Abdallah, J.~Fiaschi. S.~Khalil and S.~Moretti, in progress.
\bibitem{Khachatryan:2014fba}
CMS Collaboration,
 CMS-EXO-12-061, arXiv:1412.6302 [hep-ex].
  
\bibitem{Aad:2014cka} 
  ATLAS Collaboration,
  Phys. Rev. D {\bf 90}, 052005 (2014).

\bibitem{florianSARAH}
F.~Staub, 
Comput. Phys. Commun. {\bf 184}, 1792 (2013).

\bibitem{PorodSPheno}
W.~Porod, 
Comput. Phys. Commun. {\bf 153}, 275 (2003).

\bibitem{florianSPheno}
W.~Porod and F.~Staub, 
Comput. Phys. Commun. {\bf 183}, 2458 (2012).

\bibitem{Madgraph5}
J.~Alwall et al., JHEP {\bf 1407}, 079 (2014).

\bibitem{pythia}
  T.~Sjostrand, S.~Mrenna and P.~Z.~Skands,
  JHEP {\bf 0605}, 026 (2006).

\bibitem{PGS4}
J.~Conway,
http://www.physics.ucdavis.edu/\url{~}conway/\\research/software/pgs/pgs4-general.htm

\bibitem{Madanalysis5}
E.~Conte, B.~Fuks and G.~Serret,
Comput. Phys. Commun. {\bf 184}, 222 (2013).



\bibitem{Baer}
H.~Baer, A.~Mustafayev and X.~Tata, 
Phys. Rev. D {\bf 89}, 055007 (2014).

\bibitem{Saavedra}
C.~Han, A.~Kobakhidze, N.~Liu, A.~Saavedra,  L.~Wu and J.~M.~Yang, 
JHEP {\bf 02}, 049 (2014).


\bibitem{multijet}
ATLAS Collaboration,
  arXiv:0901.0512 [hep-ex].


\bibitem{allanach}
  B.~C.~Allanach, S.~Grab and H.~E.~Haber,
  JHEP {\bf 1101}, 138 (2011) [Erratum-ibid. {\bf 1107}, 087 (2011); Erratum-ibid. {\bf 1109}, 027 (2011)].

\bibitem{drees}
  M.~Drees, M.~Hanussek and J.~S.~Kim,
  Phys. Rev. D {\bf 86}, 035024 (2012).


\bibitem{lhc}
ATLAS Collaboration, ATLAS-CONF-2012-147; CMS Collaboration,
CMS PAS EXO-12-048.


\bibitem{atlas} 
ATLAS Collaboration, 
Phys. Rev. Lett. {\bf 110}, 011802 (2013).




\bibitem{Aad:2014vka} 
ATLAS Collaboration,
  Phys. Rev. D {\bf 90}, 012004 (2014).

\bibitem{Bell:2012rg} 
  N.~F.~Bell, J.~B.~Dent, A.~J.~Galea, T.~D.~Jacques, L.~M.~Krauss and T.~J.~Weiler,
  Phys. Rev. D {\bf 86}, 096011 (2012).



\end{thebibliography}
\end{document}